\documentclass[12pt]{article}
\usepackage{graphicx}
\usepackage{bm}

\begin{document}

\title{The two-time Leggett-Garg inequalities of a superconducting qubit interacting with thermal photons in a cavity}

\author{Hiroo Azuma\thanks{Email: zuma@nii.ac.jp}
\\
\\
{\small Global Research Center for Quantum Information Science,}\\
{\small National Institute of Informatics,}\\
{\small 2-1-2 Hitotsubashi, Chiyoda-ku, Tokyo 101-8430, Japan}\\
}

\date{\today}

\maketitle

\begin{abstract}
In this paper, we study the two-time Leggett-Garg (LG) inequalities of a quantum optical model that appears in the Josephson-junction quantum bit (qubit) interacting with an external magnetic flux.
This model is a natural extension of an exactly solvable model whose interaction between a qubit and single-mode photons is given by a product of the Pauli $z$ operator of the qubit and a linear combination of annihilation and creation operators of the photons.
By contrast, a photon's part of the interaction of our model is given by the square of the linear combination.
Because our model is not solvable, we approximately investigate its time evolution up to the second-order perturbation.
Our numerical calculations show that violation of the LG inequality diminishes as the temperature increases.
Moreover, it exhibits power laws of the temperature,
whose exponents vary depending on the coupling constant of the interaction between the qubit and photons.
The violation of the LG inequality decreases and becomes less sensitive to the temperature as the coupling constant of the interaction gets larger.
\end{abstract}

\section{\label{section-introduction}Introduction}
Several models in the field of quantum optics play essential roles in investigations of quantum information theory \cite{Alqahtani2022}.
For example, one may examine a model in which a two-level atom inside a cavity interacts with cavity mode photons with a specific wavelength.
This can be regarded as a system depicting a quantum bit (qubit) interacting with bosonic massless particles \cite{Raimond2001,Beaudoin2011,Meher2022,Bin2025}.
Furthermore, assuming that the mirrors in the cavity act as a heat bath, we can introduce the temperature for the photons.
Such models are well studied to describe the dissipation and dephasing of states of the qubit interacting with an external thermal environment \cite{Ban2020}.

From a broader perspective, atoms and photons confined within the cavity constitute an open quantum system.
One can investigate not only dissipation and depahsing but also decoherence for the qubit \cite{He2011,Kaer2014,Dajka2024}.
Moreover, by controlling the state of the photons, we can manipulate atomic states.
Reference \cite{Blais2004} proposed a method to construct the qubit using a superconducting electrical circuit based on the Jaynes-Cummings model.
Dissipation and decoherence of the multiphoton quantum Rabi oscillation in ultrastrong cavity quantum electrodynamics (QED) were investigated \cite{Garziano2015}.
Resilience of the quantum Rabi model in superconducting circuit QED was investigated \cite{Manucharyan2017}.
M\"{u}ller studied the interaction between the qubits and resonators by means of an analytical approach within the dispersive regime of the Rabi model \cite{Muller2020}.

For example, we can think of a cavity QED model with the following Hamiltonian as a typical one:
\begin{equation}
\hat{H}
=
(1/2)\hbar\Omega\hat{\sigma}_{z}
+
\hbar\omega\hat{a}^{\dagger}\hat{a}
+
\hbar\hat{\sigma}_{z}(g \hat{a}^{\dagger}+g^{*}\hat{a}),
\label{exactly-solvable-Hamiltonian-0}
\end{equation}
where $\hbar$ denotes the Planck constant,
$\Omega/(2\pi)$ denotes the frequency between the excited and ground states of the two-level atom,
$\omega/(2\pi)$ denotes the frequency of the photons confined in the cavity,
$\hat{\sigma}_{z}$ denotes the Pauli $z$ operator acting on the atom,
$\hat{a}$ and $\hat{a}^{\dagger}$ denote the annihilation and creation operators of the photons, respectively,
and $g$ denotes the coupling constant.
In early studies of quantum computers,
the noise tolerance of qubits was investigated by employing this Hamiltonian and assuming that the initial state of the photons was in thermal equilibrium obeying the Bose-Einstein distribution \cite{Palma1996,Reina2002}.
The Leggett-Garg (LG) inequalities of the system described by this Hamiltonian were investigated \cite{Usui2020}.
The reason why the model with this Hamiltonian is so popular is that it can be solved exactly.
Furthermore, this model often remains exactly solvable even when the single-mode photons are extended to multiple modes \cite{Leggett1987}.

The LG inequality is a probe that distinguishes between quantum mechanics and macroscopic local realism \cite{Leggett1985,Emary2014}.
To compute Bell's inequality, we prepare two spatially separated qubits and obtain correlation functions between their measurements.
By contrast, to evaluate the LG inequality, we examine correlations of measurements at different times for a single qubit.
Thus, the LG inequality can be regarded as a temporal analog of Bell's inequality.
Specifically, it tests the following two assumptions that are expected to hold in classical theory:
(1) macroscopic realism and (2) noninvasive measurability.

Leggett and Garg originally proposed the three-time LG inequality as a criterion for testing macroscopic local realism.
In contrast, Mawby and Halliwell introduced a two-time version of the LG inequalities \cite{Halliwell2017,Halliwell2019,Mawby2023}.
The latter are computationally more tractable, and we adopt them in this paper.
In general, the three-time LG inequality requires quantum nondemolition measurements, which makes its experimental implementation difficult.
By contrast, to evaluate the two-time LG inequalities, quantum nondemolition measurements are not necessary.
This fact is advantageous when we perform experiments.

In this paper, we consider a system whose Hamiltonian is represented by
\begin{equation}
\hat{H}
=
(1/2)\hbar\Omega\hat{\sigma}_{z}
+
\hbar\omega\hat{a}^{\dagger}\hat{a}
+
\hbar^{2}\hat{\sigma}_{z}(g \hat{a}^{\dagger}+g^{*}\hat{a})^{2},
\label{not-exactly-solvable-Hamiltonian-1}
\end{equation}
instead of the Hamiltonian given by Eq.~(\ref{exactly-solvable-Hamiltonian-0}).
We can implement the Hamiltonian of Eq.~(\ref{not-exactly-solvable-Hamiltonian-1})
by constructing a qubit in a circuit with two Josephson junctions in parallel.
We can introduce the interaction between the qubit and photons by threading a magnetic flux through the circuit.
Because we cannot rigorously solve the time evolution of the system whose Hamiltonian is given by Eq.~(\ref{not-exactly-solvable-Hamiltonian-1}),
we analyze it using the second-order perturbation theory.

Utilizing these techniques, we evaluate the two-time LG inequalities of the qubit up to the second order.
We introduce the temperature to the system by assuming that the initial states of the photons are in thermal equilibrium.
We examine the effects of the temperature on the violations of the two-time LG inequalities.
In particular, we focus on the first local minima of the violations of the LG inequalities that are smaller than zero and the times when those local minima are observed.
Regarding those values as functions of the temperature, we fit them with the power functions of the temperature approximately.
Our numerical calculations show that an increase in temperature causes the transition from quantum to classical behavior for the system.

This paper is organized as follows.
In Sec.~\ref{section-Josephson-junction-circuit}, we show how to construct the Josephson-junction qubit which interacts with thermal photons in the cavity.
In Sec.~\ref{section-two-time-Leggett-Garg}, we introduce the two-time LG inequalities and observables.
In Sec.~\ref{section-time-evolution-correlation}, we formulate the time-evolution operator and correlation functions.
In Sec.~\ref{section-computing-Tr-hat-V-hat-rho}, we give mathematical expressions that are included in the correlation functions according to the second-order perturbation theory.
In Sec.~\ref{section-numerical-calculations}, we show results of numerical calculations.
In Sec.~\ref{section-concluding-remarks}, we provide concluding remarks.
In Appendix~\ref{section-appendix-A}, we explain how to derive the mathematical expressions given in Sec.~\ref{section-computing-Tr-hat-V-hat-rho}.
In Appendix~\ref{section-appendix-B}, we show the rigorous solution of the time evolution caused by the Hamiltonian given by Eq.~(\ref{exactly-solvable-Hamiltonian-0}).

\section{\label{section-Josephson-junction-circuit}The Josephson-junction qubit interacting with the thermal photons in the cavity}
\begin{figure}
\begin{center}
\includegraphics[width=0.5\linewidth]{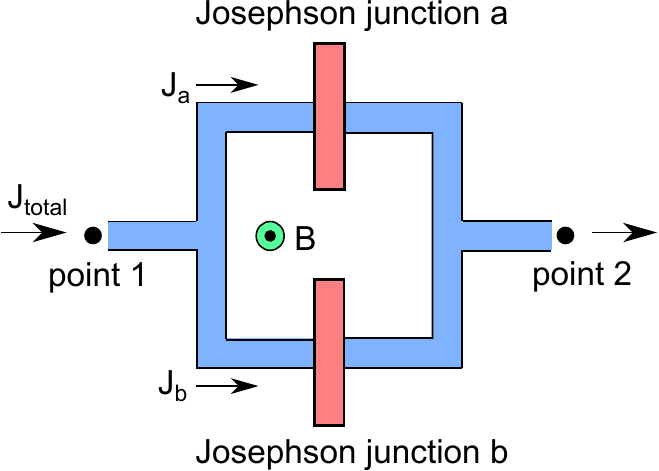}
\end{center}
\caption{A circuit with two Josephson junctions in parallel.
The magnetic field pierces the plane of the circuit.}
\label{figure01}
\end{figure}

In this section, we build the qubit with Josephson junctions, which are interacting with an external magnetic flux.
Figure~\ref{figure01} shows a circuit constructed from two Josephson junctions in parallel and a magnetic field $\bm{B}$ applied through the interior of the loop
\cite{Kittel1996,Feynman2006,Felicetti2018,Azuma2025a,Azuma2025b}.
We denote the phase differences of a wave function between points $1$ and $2$ along the paths of Josephson junctions $a$ and $b$ as $\delta_{a}$ and $\delta_{b}$, respectively.
Moreover, denoting the external magnetic flux of the magnetic field $\bm{B}$ as $\Phi$, we have
\begin{equation}
\delta_{a}=\delta_{0}-\frac{e}{\hbar c}\Phi,
\quad
\delta_{b}=\delta_{0}+\frac{e}{\hbar c}\Phi,
\end{equation}
where $\delta_{0}$ is a constant.
We denote a total current and currents along junctions $a$ and $b$ as
$J_{\mbox{\scriptsize total}}$, $J_{a}$, and $J_{b}$, respectively.
Because the Josephson effect induces $J_{a}$ and $J_{b}$,
we have
\begin{equation}
J_{\mbox{\scriptsize total}}
=
J_{a}+J_{b}
=
2J_{0}\sin\delta_{0}\cos\frac{e\Phi}{\hbar c},
\end{equation}
where $J_{0}$ is the maximum zero-voltage current.
Furthermore, because
\begin{equation}
J_{0}
\propto
\hat{\sigma}_{z}
=
|\mbox{L}\rangle\langle\mbox{L}|
-
|\mbox{R}\rangle\langle\mbox{R}|,
\end{equation}
where $|\mbox{L}\rangle$ and $|\mbox{R}\rangle$ are the left- and right-circulating persistent current states, respectively, and
\begin{equation}
\Phi
\propto
\hat{a}+\hat{a}^{\dagger},
\label{operator-flux-0}
\end{equation}
we obtain
\begin{equation}
J_{\mbox{\scriptsize total}}
\propto
\hat{\sigma}_{z}\cos(\hat{a}+\hat{a}^{\dagger})
=
\hat{\sigma}_{z}
+
\frac{1}{2}\hat{\sigma}_{z}(\hat{a}+\hat{a}^{\dagger})^{2}
+
\cdots.
\label{Jtotal-operator-expansion-0}
\end{equation}
We can write Eq.~(\ref{operator-flux-0})
because $\langle\alpha|(\hat{a}+\hat{a}^{\dagger})|\alpha\rangle=2\mbox{Re}[\alpha]$,
where the magnetic field $\bm{B}$ is represented as a coherent state $|\alpha\rangle$ with phase $\alpha$.

Here, we assume that the two Josephson junctions are placed in a cavity and the qubit is constructed from the two states $\{|\mbox{L}\rangle, |\mbox{R}\rangle\}$
interacting with single-mode photons.
Then, a total Hamiltonian of the Josephson-junction qubit and the single photons is given by Eq.~(\ref{not-exactly-solvable-Hamiltonian-1})
where
$\Omega/(2\pi)$ denotes the frequency between $|\mbox{L}\rangle$ and $|\mbox{R}\rangle$ of the Josephson-junction qubit.

From now on, we write the system of the qubit and photons as $Q$ and $P$, respectively.
We define eigenstates of $\hat{\sigma}_{z}$ as
$\hat{\sigma}_{z}|e\rangle_{Q}=|e\rangle_{Q}$
and
$\hat{\sigma}_{z}|g\rangle_{Q}=-|g\rangle_{Q}$.
We assume that mirrors surrounding the cavity can be regarded as a heat bath whose temperature is given by $(k_{\mbox{\scriptsize B}}\beta)^{-1}$,
where $k_{\mbox{\scriptsize B}}$ denotes the Boltzmann constant.
Thus, the photons obey the Bose-Einstein distribution.
Accordingly,
the initial state of the photons is in thermal equilibrium and given by
\begin{equation}
\hat{\rho}_{0,P}
=
(1-e^{-\beta\omega})
\exp(-\beta\omega\hat{a}^{\dagger}\hat{a}),
\end{equation}
where we put $\hbar=1$.
Defining the initial state of the Josephson-junction qubit as
$\hat{\rho}_{0,Q}=|g\rangle\langle g|$,
we obtain the initial state of the whole system
as
$\hat{\rho}_{0}=\hat{\rho}_{0,Q}\otimes\hat{\rho}_{0,P}$.

\section{\label{section-two-time-Leggett-Garg}The two-time LG inequalities and observables}
In this section, we introduce the two-time LG inequalities and observables for the qubit.
The two-time LG inequalities are defined as
\begin{equation}
\mbox{LG}_{s_{0},s_{1}}
=
\langle(1+s_{0}\hat{A}_{0})(1+s_{1}\hat{A}_{1})\rangle\geq 0,
\label{two-time-LG-0}
\end{equation}
\begin{equation}
\langle \hat{A}_{0}\rangle
=
\sum_{A_{0}\in\{\pm 1\}}p(A_{0})A_{0},
\label{expectation-value-A0}
\end{equation}
\begin{equation}
\langle \hat{A}_{1}\rangle
=
\sum_{A_{1}\in\{\pm 1\}}p(A_{1})A_{1},
\label{expectation-value-A1}
\end{equation}
\begin{equation}
\langle \hat{A}_{0}\hat{A}_{1}\rangle
=
\sum_{A_{0},A_{1}\in\{\pm 1\}}p(A_{0},A_{1})A_{0}A_{1},
\label{correlation-A0A1}
\end{equation}
where $s_{0}, s_{1}\in\{\pm 1\}$,
$\hat{A}_{0}$ and $\hat{A}_{1}$ are observables of the qubit at time $t_{0}$ and $t_{1}$,
respectively,
$A_{0}$ and $A_{1}$ are values of outcomes of measurements with $\hat{A}_{0}$ and $\hat{A}_{1}$, respectively,
$p(A_{0},A_{1})$ is a probability that $A_{0}$ and $A_{1}$ are obtained,
$p(A_{0})=\sum_{A_{1}\in\{\pm 1\}}p(A_{0},A_{1})$,
and
$p(A_{1})=\sum_{A_{0}\in\{\pm 1\}}p(A_{0},A_{1})$.
Here, we set $t_{0}=0$, that is, the initial time, and $t_{1}>0$.

Next, we define the observable of the qubit as
\begin{equation}
\hat{A}
=
a_{x}\hat{\sigma}_{x}
+
a_{y}\hat{\sigma}_{y}
+
a_{z}\hat{\sigma}_{z}
=
a\hat{\sigma}_{+}
+
a^{*}\hat{\sigma}_{-}
+
a_{z}\hat{\sigma}_{z},
\label{A-observable-explicit-0}
\end{equation}
where
$\hat{\sigma}_{\pm}=(1/2)(\hat{\sigma}_{x}\pm i \hat{\sigma}_{y})$,
$a=a_{x}+i a_{y}$,
and $\hat{\sigma}_{x}$, $\hat{\sigma}_{y}$, and $\hat{\sigma}_{z}$
are Pauli matrices acting on the qubit.
Because a term which is proportional to the identity matrix $\hat{\bm{I}}$
only adds a constant to the outcomes of the measurements and it does not affect physics essentially,
we drop $\hat{\bm{I}}$ in Eq.~(\ref{A-observable-explicit-0}).
Moreover, assuming that eigenvalues of $\hat{A}$ is given by $\pm 1$,
we obtain
\begin{equation}
|a_{x}|^{2}+|a_{y}|^{2}+|a_{z}|^{2}=|a|^{2}+|a_{z}|^{2}=1.
\label{a-constants-normalization}
\end{equation}
Projection operators $\{\hat{\Pi}_{\mu}: \mu=\pm 1\}$ are given by
\begin{equation}
\hat{\Pi}_{\mu}
=
(1/2)
(\hat{\bm{I}}+\mu\hat{A}).
\label{definition-projections-0}
\end{equation}

\section{\label{section-time-evolution-correlation}The time-evolution operator and correlation functions}
In this section, we formulate a time-evolution unitary operator and correlation functions.
First of all, we construct the time-evolution operator with the Hamiltonian given by Eq.~(\ref{not-exactly-solvable-Hamiltonian-1}) as
\begin{eqnarray}
\hat{U}(t_{1},t_{0})
&=&
\exp[-i\Omega(t_{1}-t_{0})/2]
\exp[-i\hat{H}_{+}(t_{1}-t_{0})]
|e\rangle\langle e| \nonumber \\
&&
+
\exp[i\Omega(t_{1}-t_{0})/2]
\exp[-i\hat{H}_{-}(t_{1}-t_{0})]
|g\rangle\langle g|,
\end{eqnarray}
where
\begin{equation}
\hat{H}_{\pm}
=
\omega\hat{a}^{\dagger}\hat{a}
\pm
(g\hat{a}^{\dagger}+g^{*}\hat{a})^{2},
\end{equation}
$t_{1}>t_{0}$,
and $\hat{U}(t_{1},t_{0})$ transform a state at time $t_{0}$ into a state at time $t_{1}$.
Here, we rewrite $\hat{H}_{\pm}$ as the following convenient forms:
\begin{equation}
\hat{H}_{\pm}
=
\omega_{\pm}\hat{a}^{\dagger}\hat{a}
\pm
(\kappa\hat{a}^{2}+\kappa^{*}\hat{a}^{2})
\pm
|g|^{2},
\end{equation}
where $\omega_{\pm}=\omega\pm 2|g|^{2}$ and $\kappa=g^{2}$.
Next, we compute a probability $P(\mu,t_{1};\nu,t_{0})$
that we observe $\mu$ and $\nu$ at time $t_{1}$ and time $t_{0}$, respectively,
where $\mu,\nu\in\{\pm 1\}$,
using the projection operators $\{\hat{\Pi}_{\pm 1}\}$
defined by Eq.~(\ref{definition-projections-0}).
Because it is given in the form,
\begin{equation}
P(\mu,t_{1};\nu,t_{0})
=
\mbox{Tr}[\hat{\Pi}_{\nu}(0)\hat{\Pi}_{\mu}(t_{1}-t_{0})\hat{\Pi}_{\nu}(0)\hat{\rho}_{0}],
\end{equation}
where
\begin{equation}
\hat{\Pi}_{\mu}(t_{1}-t_{0})
=
\hat{U}^{\dagger}(t_{1},t_{0})
\hat{\Pi}_{\mu}
\hat{U}(t_{1},t_{0}),
\label{definition-hat-Pi-mu-t1-t0}
\end{equation}
and we define
\begin{eqnarray}
\hat{A}(t_{1}-t_{0})
&=&
\hat{U}^{\dagger}(t_{1},t_{0})
\hat{A}
\hat{U}(t_{1},t_{0}) \nonumber \\
&=&
\exp[i\Omega(t_{1}-t_{0})]
\hat{V}(t_{1},t_{0})
a\hat{\sigma}_{+}
+
\exp[-i\Omega(t_{1}-t_{0})]
\hat{V}^{\dagger}(t_{1},t_{0})
a^{*}\hat{\sigma}_{-} \nonumber \\
&&
+
a_{z}\hat{\sigma}_{z},
\label{definition-hat-A-t1-t0}
\end{eqnarray}
where
\begin{equation}
\hat{V}(t_{1},t_{0})
=
\exp[i\hat{H}_{+}(t_{1}-t_{0})]
\exp[-i\hat{H}_{-}(t_{1}-t_{0})],
\label{definition-hat-V-t1-t0}
\end{equation}
we obtain
\begin{eqnarray}
P(\mu,t_{1};\nu,t_{0})
&=&
\frac{3}{8}
+
\frac{1}{4}\nu\langle \hat{A}(0)\rangle_{0}
+
\frac{1}{8}\mu\langle \hat{A}(t_{1}-t_{0})\rangle_{0}
+
\frac{1}{4}\mu\nu\mbox{Re}[\langle \hat{A}(t_{1}-t_{0})\hat{A}(0)\rangle_{0}] \nonumber\\
&&
+
\frac{1}{8}\mu\langle \hat{A}(0)\hat{A}(t_{1}-t_{0})\hat{A}(0)\rangle_{0},
\label{P-mu-t1-nu-t0}
\end{eqnarray}
where
$\langle \hat{X}\rangle_{0}=\mbox{Tr}[\hat{X}\hat{\rho}_{0}]$.
Here, we draw attention that $p(A_{0},A_{1})=P(A_{1},t_{1};A_{0},t_{0})$.

We can calculate
$\langle \hat{A}_{0}\rangle=\langle \hat{A}(0)\rangle_{0}$
given by Eq.~(\ref{expectation-value-A0})
with ease because
\\
$\langle \hat{A}\rangle_{0}=\mbox{Tr}[\hat{A}(0)\hat{\rho}_{0,Q}]$.
According to Eqs.~(\ref{expectation-value-A1}), (\ref{correlation-A0A1}),
and (\ref{P-mu-t1-nu-t0}),
we obtain
\begin{equation}
\langle \hat{A}_{1}\rangle
=
\frac{1}{2}
\langle \hat{A}(t_{1}-t_{0})\rangle_{0}
+
\frac{1}{2}
\langle \hat{A}(0)\hat{A}(t_{1}-t_{0})\hat{A}(0)\rangle_{0},
\label{expectation-value-hat-A1}
\end{equation}
\begin{equation}
\langle \hat{A}_{0}\hat{A}_{1}\rangle
=
\mbox{Re}[\langle \hat{A}(t_{1}-t_{0})\hat{A}(0)\rangle_{0}].
\label{expectation-value-hat-A0-hat-A1}
\end{equation}
Using Eq.~(\ref{definition-hat-A-t1-t0}),
we can rewrite $\langle \hat{A}(t_{1}-t_{0})\rangle_{0}$,
$\langle \hat{A}(t_{1}-t_{0})\hat{A}(0)\rangle_{0}$,
and $\langle \hat{A}(0)\hat{A}(t_{1}-t_{0})\hat{A}(0)\rangle_{0}$ as
\begin{equation}
\langle \hat{A}(t_{1}-t_{0})\rangle_{0}
=
2\mbox{Re}
\{
\exp[i\Omega(t_{1}-t_{0})]
a
\mbox{Tr}[\hat{\sigma}_{+}\hat{\rho}_{0,Q}]
\mbox{Tr}[\hat{V}(t_{1},t_{0})\hat{\rho}_{0,P}]
\}
+
a_{z}\mbox{Tr}[\hat{\sigma}_{z}\hat{\rho}_{0,Q}],
\label{expectation-value-hat-A-t1-t0-0}
\end{equation}
\begin{eqnarray}
\langle \hat{A}(t_{1}-t_{0})\hat{A}(0)\rangle_{0}
&=&
2\mbox{Re}
\{
\exp[i\Omega(t_{1}-t_{0})]
a
\mbox{Tr}[\hat{\sigma}_{+}\hat{A}(0)\hat{\rho}_{0,Q}]
\mbox{Tr}[\hat{V}(t_{1},t_{0})\hat{\rho}_{0,P}]
\} \nonumber \\
&&
+
a_{z}
\mbox{Tr}[\hat{\sigma}_{z}\hat{A}(0)\hat{\rho}_{0,Q}],
\label{expectation-value-hat-A-t1-to-hat-A}
\end{eqnarray}
\begin{eqnarray}
\langle \hat{A}(0)\hat{A}(t_{1}-t_{0})\hat{A}(0)\rangle_{0}
&=&
2\mbox{Re}
\{
\exp[i\Omega(t_{1}-t_{0})]
a
\mbox{Tr}[\hat{A}(0)\hat{\sigma}_{+}\hat{A}(0)\hat{\rho}_{0,Q}]
\mbox{Tr}[\hat{V}(t_{1},t_{0})\hat{\rho}_{0,P}]
\} \nonumber \\
&&
+
a_{z}
\mbox{Tr}[\hat{A}(0)\hat{\sigma}_{z}\hat{A}(0)\hat{\rho}_{0,Q}].
\label{expectation-value-hat-A-hat-A-t1-to-hat-A}
\end{eqnarray}
Because of Eqs.~(\ref{expectation-value-hat-A-t1-t0-0}), (\ref{expectation-value-hat-A-t1-to-hat-A}), and (\ref{expectation-value-hat-A-hat-A-t1-to-hat-A}),
we understand that what we must do is computing $\mbox{Tr}[\hat{V}(t_{1},t_{0})\hat{\rho}_{0,P}]$.

\section{\label{section-computing-Tr-hat-V-hat-rho}Evaluation of $\mbox{Tr}[\hat{V}(t_{1},t_{0})\hat{\rho}_{0,P}]$}
As explained in Appendix~\ref{section-appendix-A},
$\mbox{Tr}[\hat{V}(t_{1},t_{0})\hat{\rho}_{0,P}]$ is given in the form,
\begin{equation}
\mbox{Tr}[\hat{V}(t_{1},t_{0})\hat{\rho}_{0,P}]
=
(1-e^{-\beta\omega})
\sum_{m=0}^{\infty}
\exp(-m\beta\omega){\cal F}(m),
\label{Tr-hat-V-t1-t0-hat-rho-0-P}
\end{equation}
where
\begin{eqnarray}
{\cal F}(m)
&=&
F(m)
\left\{
1-\frac{1}{2}|\tilde{\kappa}_{+}|^{2}
[\theta(m-2)m(m-1)+(m+1)(m+2)]
\right\} \nonumber \\
&&
\times
\left\{
1-\frac{1}{2}|\tilde{\kappa}_{-}|^{2}
[\theta(m-2)m(m-1)+(m+1)(m+2)]
\right\} \nonumber \\
&&
-F(m+2)\tilde{\kappa}_{+}^{*}\tilde{\kappa}_{-}(m+1)(m+2)
-F(m-2)\tilde{\kappa}_{-}^{*}\tilde{\kappa}_{+}\theta(m-2)m(m-1) \nonumber \\
&&
+
\frac{1}{4}F(m+4)\tilde{\kappa}_{+}^{* 2}\tilde{\kappa}_{-}^{2}(m+1)(m+2)(m+3)(m+4) \nonumber \\
&&
+
\frac{1}{4}F(m-4)\tilde{\kappa}_{-}^{* 2}\tilde{\kappa}_{+}^{2}\theta(m-4)m(m-1)(m-2)(m-3),
\label{cal-F}
\end{eqnarray}
\begin{eqnarray}
F(n)
&=&
\exp
\left[
iT^{2}|\kappa|^{2}
\frac{\tilde{T}_{+}-\sin\tilde{T}_{+}}{\tilde{T}_{+}}
f(n)
\right]
e^{i H_{0}^{(+)}T}
e^{-i H_{0}^{(-)}T} \nonumber \\
&&
\times
\exp
\left[
-iT^{2}|\kappa|^{2}
\frac{\tilde{T}_{-}-\sin\tilde{T}_{-}}{\tilde{T}_{-}}
f(n)
\right]
+
O(|\kappa|^{3}),
\end{eqnarray}
\begin{equation}
\tilde{\kappa}_{\pm}
=
\frac{T}{\tilde{T}_{\pm}}
(e^{i\tilde{T}_{\pm}}-1)\kappa,
\label{tilde-kappa-plus-minus-definition-0}
\end{equation}
\begin{equation}
\theta(m)
=
\left\{
\begin{array}{cl}
1 & m\geq 0 \\
0 & m<0 \\
\end{array}
\right.,
\end{equation}
$\tilde{T}_{\pm}=2T\omega_{\pm}$,
$T=t_{1}-t_{0}$,
$H_{0}^{(\pm)}=\omega_{\pm}n\pm |g|^{2}$,
and $f(n)=-2(1+2n)$.
Here, we draw attention to the fact that $T=t_{1}-t_{0}$ is different from the temperature $(k_{\mbox{\scriptsize B}}\beta)^{-1}$.
Furthermore, $\tilde{\kappa}_{\pm}$ and $\tilde{T}_{\pm}$ are dimensionless quantities.

\section{\label{section-numerical-calculations}Numerical calculations}
\begin{figure}
\begin{center}
\includegraphics[width=0.8\linewidth]{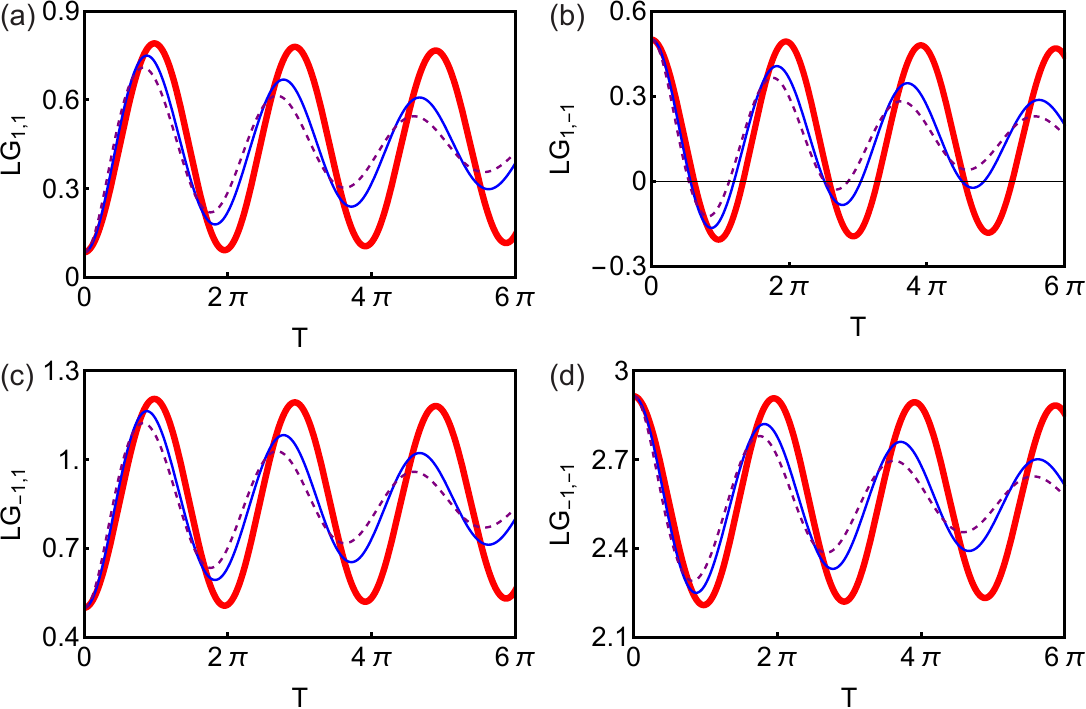}
\end{center}
\caption{Plots of $\mbox{LG}_{s_{0},s_{1}}$ for $s_{0},s_{1}\in\{\pm 1\}$ as functions of
$T=t_{1}-t_{0}(\geq 0)$.
Curves of (a), (b), (c), and (d) represent $\mbox{LG}_{1,1}$, $\mbox{LG}_{1,-1}$, $\mbox{LG}_{-1,1}$, and $\mbox{LG}_{-1,-1}$, respectively.
In those plots, we set $\Omega=1$, $g=0.075$, $\omega=0.1$, $a_{x}=a_{z}=1/\sqrt{2}$, and $a_{y}=0$.
The thick solid red, thin solid blue, and thin dashed purple curves represent $\beta=10, 1.5$, and $0.8$, respectively, for (a), (b), (c), and (d).
Looking at (b), we note that $\mbox{LG}_{1,-1}<0$ at specific times, so that violation of the two-time LG inequality occurs at those times.}
\label{figure02}
\end{figure}

In Fig.~\ref{figure02}, we plot $\mbox{LG}_{s_{0},s_{1}}$ for $s_{0},s_{1}\in\{\pm 1\}$
given by Eq.~(\ref{two-time-LG-0})
as functions of time $T=t_{1}-t_{0}(\geq 0)$.
Looking at Figs.~\ref{figure02}(a), (b), (c), and (d),
we note that only $\mbox{LG}_{1,-1}$ violates the two-time LG inequality.
Comparing the curves of $\beta=10, 1.5$, and $0.8$ in Fig.~\ref{figure02}(b),
the first local minimum value of $\mbox{LG}_{1,-1}$ from the time $T=0$ becomes larger as the temperature increases.
Simultaneously, the time when $\mbox{LG}_{1,-1}$ takes the first local minimum value approaches zero as the temperature increases.
Hence, we concentrate on calculating $\mbox{LG}_{1,-1}$.

\begin{figure}
\begin{center}
\includegraphics[width=0.8\linewidth]{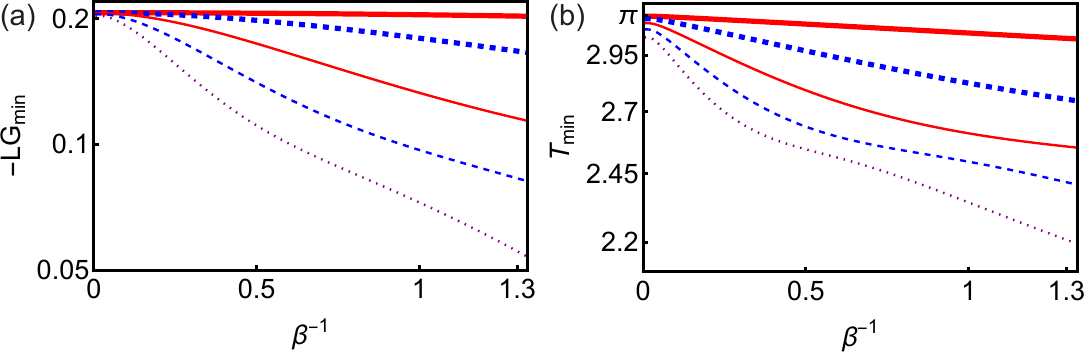}
\end{center}
\caption{(a) Plots of $-\mbox{LG}_{\mbox{\scriptsize min}}$ as a function of the temperature $\beta^{-1}$, where $\mbox{LG}_{\mbox{\scriptsize min}}$ represents the first local minimum value of $\mbox{LG}_{1,-1}$ as a function of the time $T=t_{1}-t_{0}(\geq 0)$.
(b) Plots of $T_{\mbox{\scriptsize min}}$ as a function of the temperature $\beta^{-1}$, where $T_{\mbox{\scriptsize min}}$ represents the time when $\mbox{LG}_{1,-1}$ takes the first local minimum value $\mbox{LG}_{\mbox{\scriptsize min}}$.
The thick solid red, thick dashed blue, thin solid red, thin dashed blue, and thin dotted purple curves represent $g=0.26, 0.52, 0.78, 1.04$, and $1.3$, respectively.
The other parameters are set as $\Omega=1$, $\omega=0.1$, $a_{x}=a_{z}=1/\sqrt{2}$, and $a_{y}=0$.
In graphs (a) and (b), the vertical axis is shown on a logarithmic scale.}
\label{figure03}
\end{figure}

In Fig.~\ref{figure03},
we plot $-\mbox{LG}_{\mbox{\scriptsize min}}$ and $T_{\mbox{\scriptsize min}}$ as functions of the temperature $\beta^{-1}$,
where $\mbox{LG}_{\mbox{\scriptsize min}}$ represents the first local minimum value of $\mbox{LG}_{1,-1}$ from $T=0$
and $T_{\mbox{\scriptsize min}}$ represents the time when $\mbox{LG}_{1,-1}$ take the value $\mbox{LG}_{\mbox{\scriptsize min}}$.
As the temperature increases, $-\mbox{LG}_{\mbox{\scriptsize min}}$ and $T_{\mbox{\scriptsize min}}$ get smaller.
This implies that the violation of the two-time LG inequality diminishes and the activity of the system is gradually governed by classical physics as the temperature increases.
At the same time, the violation of the two-time LG inequality is reduced as the coupling constant $g$ gets larger.
We can interpret this fact as follows.
As the coupling constant of the interaction between the qubit and photons in the cavity increases, the effect of the heat bath on the qubit via the cavity photons enlarges.
Thus, the increase of $g$ lets the violation of the LG inequality decrease.

\begin{figure}
\begin{center}
\includegraphics[width=0.8\linewidth]{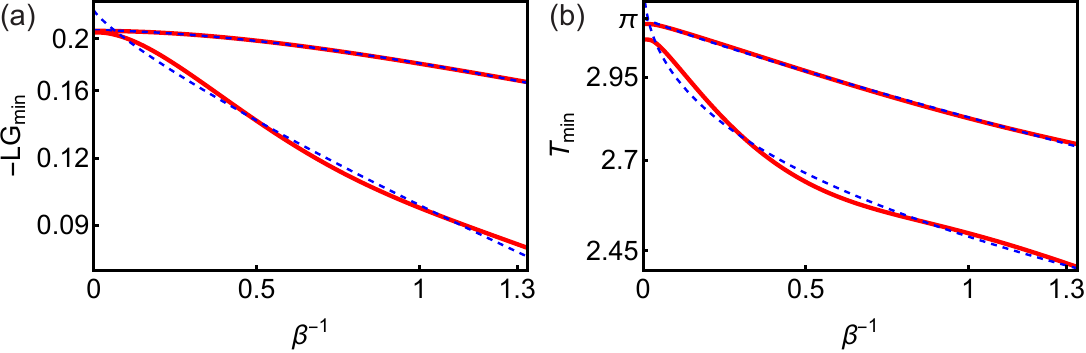}
\end{center}
\caption{(a) The solid red curves represent plots of $-\mbox{LG}_{\mbox{\scriptsize min}}$ as functions of the temperature $\beta^{-1}$.
The dashed blue curves represent the results of fitting approximate functions
$a_{1}\beta^{-b_{1}}+c_{1}$ to $\log(-\mbox{LG}_{\mbox{\scriptsize min}})$.
(b) The solid red curves represent plots of $T_{\mbox{\scriptsize min}}$ as functions of the temperature $\beta^{-1}$.
The dashed blue curves represent the results of fitting
approximate functions
$a_{2}\beta^{-b_{2}}+c_{2}$ to $\log T_{\mbox{\scriptsize min}}$.
For both graphs (a) and (b),
the upper solid red and dashed blue curves correspond to the case of $g=0.52$,
while the lower solid red and dashed blue curves correspond to the case of $g=1.04$.
In the graphs, the vertical axis is shown on a logarithmic scale.
}
\label{figure04}
\end{figure}

In Figs.~\ref{figure04}(a) and (b), we perform curve fitting to the data of
$\{(\beta^{-1},\log(-\mbox{LG}_{\mbox{\scriptsize min}}))\}$
and
$\{(\beta^{-1},\log T_{\mbox{\scriptsize min}})\}$
using the approximate functions,
\begin{equation}
\log(-\mbox{LG}_{\mbox{\scriptsize min}})
=
a_{1}\beta^{-b_{1}}+c_{1},
\label{fitting-minus-LG-min}
\end{equation}
and
\begin{equation}
\log T_{\mbox{\scriptsize min}}
=
a_{2}\beta^{-b_{2}}+c_{2},
\label{fitting-T-min}
\end{equation}
respectively.
Looking at Figs.~\ref{figure04}(a) and (b), we note that the curve fittings are reliable for $g=0.52$ but unreliable for $g=1.04$.
Thus, we apply these curve fittings to the data for $0<g\leq 0.52$.

\begin{figure}
\begin{center}
\includegraphics[width=0.8\linewidth]{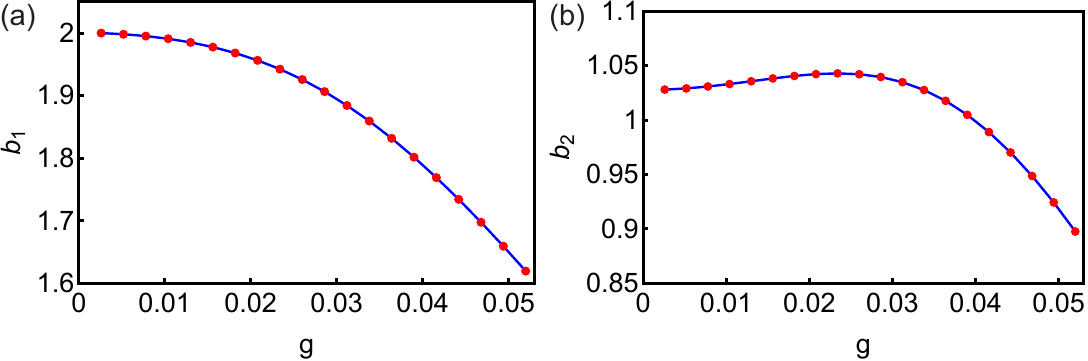}
\end{center}
\caption{(a) A plot of the constant $b_{1}$ as a function of $g$, where $b_{1}$ is obtained by the fitting of $\{(\beta^{-1},\log(-\mbox{LG}_{\mbox{\scriptsize min}}))\}$
with the approximate function $a_{1}\beta^{-b_{1}}+c_{1}$.
(b) A plot of the constant $b_{2}$ as a function of $g$, where $b_{2}$ is obtained by the fitting of $\{(\beta^{-1},\log T_{\mbox{\scriptsize min}})\}$
with the approximate function $a_{2}\beta^{-b_{2}}+c_{2}$.
In both graphs (a) and (b), the red points represent the data obtained from the fitting.
We join those points with blue curves.}
\label{figure05}
\end{figure}

In Figs.~\ref{figure05}(a) and (b), we plot the constants $b_{1}$ and $b_{2}$ as functions of $g$, respectively.
Figure~\ref{figure05}(a) shows that $b_{1}$ decreases as $g$ gets larger.
Figure~\ref{figure05}(b) shows that $b_{2}$ has a local maximum at $g=0.0234$.
Because we fit the data using the functions given by Eqs.~(\ref{fitting-minus-LG-min}) and (\ref{fitting-T-min}),
$-\mbox{LG}_{\mbox{\scriptsize min}}$ and $T_{\mbox{\scriptsize min}}$ becomes more sensitive to the change of the temperature
if $b_{1}$ and $b_{2}$ gets larger.
The exponent $b_{1}$ monotonically decreases as the coupling constant $g$ gets larger.
Although the exponent $b_{2}$ takes the local maximum at $g=0.0234$, it monotonically decreases as $g$ gets larger for $g>0.0234$.
Hence, the $\mbox{LG}_{1,-1}$ becomes less sensitive to the temperature as $g$ gets larger.
We note that the increase in the coupling constant $g$ makes the violation of the LG inequality smaller in Fig.~\ref{figure03}.
At the same time, the sensitivity of the violation to temperature is reduced as $g$ increases.

\begin{figure}
\begin{center}
\includegraphics[width=0.5\linewidth]{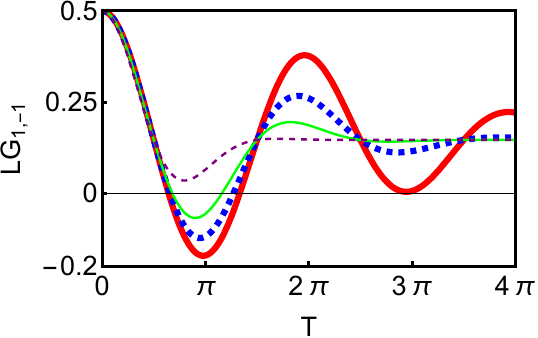}
\end{center}
\caption{Plots of $\mbox{LG}_{1,-1}$ of the system whose Hamiltonian is given by Eq.~(\ref{exactly-solvable-Hamiltonian-0}) as functions of the time $T$.
The thick solid red, thick dashed blue, thin solid green, and thin dashed purple curves represent $\beta=\infty, 8, 4$, and $1.5$, respectively.
In those plots, we set $\Omega=1$, $g=0.075$, $\omega=0.1$, $a_{x}=a_{z}=1/\sqrt{2}$, and $a_{y}=0$.
Looking at the thick solid red curve, that is, the zero-temperature case, we note that the first local minimum at $T=\pi$ violates the LG inequality but the second local minimum at $T=3\pi$ does not violate it.}
\label{figure06}
\end{figure}

Finally, we compare the LG inequalities of the Hamiltonian given by Eq.~(\ref{not-exactly-solvable-Hamiltonian-1}) with those of the Hamiltonian given by Eq.~(\ref{exactly-solvable-Hamiltonian-0}).
In the system governed by the Hamiltonian of Eq.~(\ref{exactly-solvable-Hamiltonian-0}),
$\mbox{LG}_{1,-1}$ violates the LG inequality but $\mbox{LG}_{1,1}$, $\mbox{LG}_{-1,1}$, and $\mbox{LG}_{-1,-1}$ obey the LG inequalities as in the cases for the system whose Hamiltonian is given by Eq.~(\ref{not-exactly-solvable-Hamiltonian-1}).
In Fig.~\ref{figure06}, we plot $\mbox{LG}_{1,-1}$ as a function of the time $T$ for the inverses of temperature $\beta=\infty, 8, 4$, and $1.5$.
Looking at Fig.~\ref{figure06}, we note that the curve of $\beta=1.5$ does not violate the LG inequality.
By contrast, Fig.~\ref{figure02}(b) shows that the curve of $\beta=1.5$ violates the LG inequality.
Thus, we conclude that the system whose Hamiltonian is given by Eq.~(\ref{exactly-solvable-Hamiltonian-0}) is less robust than that with the Hamiltonian of Eq.~(\ref{not-exactly-solvable-Hamiltonian-1}) from the viewpoint of quantumness.

\section{\label{section-concluding-remarks}Concluding remarks}
In this paper, we derived a quantum optical model implemented with a superconducting Josephson-junction circuit and investigated its time evolution
with initial states of the photons in the thermal equilibrium.
Because we were not able to solve this model exactly,
we used the second-order perturbation theory.
Calculating correlation functions, we evaluated the two-time LG inequalities and found their violations.
By calculating the first local minimum values of the LG inequality that were smaller than zero and the times when those values were observed, we plotted them as functions of the temperature, fitted them using power functions of the temperature, and obtained their exponents.
Our numerical estimations showed that the violation of the LG inequality decreases as the temperature increases.
Moreover, the violation of the LG inequality becomes smaller and less sensitive to temperature as the coupling constant increases.

A variety of quantum optical models have been proposed to study quantum information processing, including the multiphoton Jaynes-Cummings model \cite{Azuma2025a,Azuma2025b,Tang2023,Azuma2024}.
Further progress in quantum information processing is expected along this line.

\section*{Acknowledgment}
This work was supported by the MEXT Quantum Leap Flagship Program Grant No. JPMXS0120351339.

\appendix

\section{\label{section-appendix-A}Derivation of $\mbox{Tr}[\hat{V}(t_{1},t_{0})\hat{\rho}_{0,P}]$}
In this section, we compute $\mbox{Tr}[\hat{V}(t_{1},t_{0})\hat{\rho}_{0,P}]$ defined in Sec.~\ref{section-time-evolution-correlation} with the second-order perturbation theory.
Because of Eq.~(\ref{definition-hat-V-t1-t0}), what we must do is to evaluate
$\exp[i\hat{H}_{+}(t_{1}-t_{0})]
\exp[-i\hat{H}_{-}(t_{1}-t_{0})]$.
Defining
$\hat{H}_{\pm}=\hat{H}_{0}^{(\pm)}\pm \hat{V}$,
$\hat{H}_{0}^{(\pm)}=\omega_{\pm}\hat{a}^{\dagger}\hat{a}\pm|g|^{2}$,
$\hat{V}=\kappa\hat{a}^{\dagger 2}+\kappa^{*}\hat{a}^{2}$,
and $\Delta t=T/N$,
we can decompose $\exp[i\hat{H}_{+}(t_{1}-t_{0})]=\exp(i\hat{H}_{+}T)$
by the Lie-Trotter product formula as follows:
\begin{eqnarray}
\exp(i\hat{H}_{+}T)
&=&
\lim_{N\to\infty}
(e^{i\hat{H}_{0}^{(+)}\Delta t} e^{i\hat{V}\Delta t})^{N} \nonumber \\
&=&
\lim_{N\to\infty}
(e^{i\hat{H}_{0}^{(+)}\Delta t} e^{i\hat{V}\Delta t}e^{-i\hat{H}_{0}^{(+)}\Delta t})
(e^{2i\hat{H}_{0}^{(+)}\Delta t} e^{i\hat{V}\Delta t}e^{-2i\hat{H}_{0}^{(+)}\Delta t}) \nonumber \\
&&
\times\cdots
\times
(e^{Ni\hat{H}_{0}^{(+)}\Delta t} e^{i\hat{V}\Delta t}e^{-Ni\hat{H}_{0}^{(+)}\Delta t})
e^{Ni\hat{H}_{0}^{(+)}\Delta t} \nonumber \\
&=&
\lim_{N\to\infty}
\exp
[i\Delta t(e^{2i\Delta t\omega_{+}}\kappa\hat{a}^{\dagger 2}+e^{-2i\Delta t\omega_{+}}\kappa^{*}\hat{a}^{2})] \nonumber \\
&&
\times
\exp
[i\Delta t(e^{4i\Delta t\omega_{+}}\kappa\hat{a}^{\dagger 2}+e^{-4i\Delta t\omega_{+}}\kappa^{*}\hat{a}^{2})] \nonumber \\
&&
\times\cdots
\times
\exp
[i\Delta t(e^{2Ni\Delta t\omega_{+}}\kappa\hat{a}^{\dagger 2}+e^{-2Ni\Delta t\omega_{+}}\kappa^{*}\hat{a}^{2})]
e^{i\hat{H}_{0}^{(+)}T}.
\label{exp-i-hat-H-plut-T-production-0}
\end{eqnarray}
Here, we evaluate $e^{\hat{A}}e^{\hat{B}}$, where
\begin{eqnarray}
\hat{A}
&=&
i\Delta t(c\hat{a}^{\dagger 2}+c^{*}\hat{a}^{2}), \nonumber \\
\hat{B}
&=&
i\Delta t(c'\hat{a}^{\dagger 2}+c'^{*}\hat{a}^{2}),
\end{eqnarray}
$c=e^{2i\Delta t\omega_{+}}\kappa$, and $c'=e^{4i\Delta t\omega_{+}}\kappa$,
using the Baker-Campbell-Hausdorff formula,
\begin{equation}
e^{\hat{A}}e^{\hat{B}}
=
\exp[\hat{A}+\hat{B}+\frac{1}{2}[\hat{A},\hat{B}]+\frac{1}{12}[\hat{A}-\hat{B},[\hat{A},\hat{B}]]+\cdots].
\end{equation}
Accordingly, we obtain
\begin{equation}
e^{\hat{A}}e^{\hat{B}}
=
\exp[\hat{A}+\hat{B}+i(\Delta t)^{2}\sin(2\Delta t\omega_{+})|\kappa|^{2}f(\hat{n})+O(|\kappa|^{3})],
\end{equation}
where $f(\hat{n})=[\hat{a}^{\dagger 2},\hat{a}^{2}]=-2(1+2\hat{n})$ and $\hat{n}=\hat{a}^{\dagger}\hat{a}$.
To obtain $e^{i\hat{H}_{+}T}$ given by Eq.~(\ref{exp-i-hat-H-plut-T-production-0})
up to the second order of $\kappa$,
we must compute a term on the order of $(\Delta t)^{2}$ that appears in the exponent.
If we define
\begin{equation}
\hat{C}
=
i\Delta t(c''\hat{a}^{\dagger 2}+c''^{*}\hat{a}^{2}),
\end{equation}
where $c''=e^{6i\Delta t\omega_{+}}\kappa$,
we obtain
\begin{equation}
[\hat{A}+\hat{B},\hat{C}]
=
2i(\Delta t)^{2}[\sin(2\Delta t\omega_{+})+\sin(4\Delta t\omega_{+})]|\kappa|^{2}f(\hat{n}).
\end{equation}
Hence, the term on the order of $(\Delta t)^{2}$ that appears in the exponent of $e^{i\hat{H}_{+}T}$ is given by
\begin{equation}
2i(\Delta t)^{2}|\kappa|^{2}f(\hat{n})
\{
(N-1)\sin(2\Delta t\omega_{+})
+
(N-2)\sin(4\Delta t\omega_{+})
+
\cdots
+
\sin[2(N-1)\Delta t\omega_{+}]
\}.
\end{equation}
Because
\begin{equation}
\lim_{N\to\infty}\frac{1}{N^{2}}
\sum_{k=1}^{N-1}(N-k)\sin(k\epsilon_{+})
=
\frac{\tilde{T}_{+}-\sin\tilde{T}_{+}}{\tilde{T}_{+}},
\end{equation}
and
\begin{equation}
\lim_{N\to\infty}
\frac{1}{N}
\sum_{k=1}^{N}
e^{\pm ik\epsilon_{+}}
=
\mp\frac{i}{\tilde{T}_{+}}(e^{\pm i\tilde{T}_{+}}-1),
\end{equation}
where $\epsilon_{+}=2\Delta t\omega_{+}$ and $\tilde{T}_{+}=N\epsilon_{+}=2T\omega_{+}$,
we attain
\begin{eqnarray}
e^{i\hat{H}_{+}T}
&=&
\exp
\left[
\frac{T}{\tilde{T}_{+}}
(e^{i\tilde{T}_{+}}-1)\kappa\hat{a}^{\dagger 2}
-
\frac{T}{\tilde{T}_{+}}
(e^{-i\tilde{T}_{+}}-1)\kappa^{*}\hat{a}^{2}
\right] \nonumber \\
&&
\times
\exp
\left[
iT^{2}|\kappa|^{2}
\frac{\tilde{T}_{+}-\sin\tilde{T}_{+}}{\tilde{T}_{+}}f(\hat{n})
\right]
e^{i\hat{H}_{0}^{(+)}T}
+
O(|\kappa|^{3}).
\end{eqnarray}
In the same way as above, we obtain
\begin{eqnarray}
e^{-i\hat{H}_{-}T}
&=&
e^{-i\hat{H}_{0}^{(-)}T}
\exp
\left[
-iT^{2}|\kappa|^{2}
\frac{\tilde{T}_{-}-\sin\tilde{T}_{-}}{\tilde{T}_{-}}f(\hat{n})
\right] \nonumber \\
&&
\times
\exp
\left[
\frac{T}{\tilde{T}_{-}}
(e^{i\tilde{T}_{-}}-1)\kappa\hat{a}^{\dagger 2}
-
\frac{T}{\tilde{T}_{-}}
(e^{-i\tilde{T}_{-}}-1)\kappa^{*}\hat{a}^{2}
\right]
+
O(|\kappa|^{3}).
\end{eqnarray}

Next, we calculate a matrix element
\begin{equation}
{}_{P}\langle m|
\hat{V}(t_{1},t_{0})
|m\rangle_{P}
=
{}_{P}\langle m|
\exp(\tilde{\kappa}_{+}\hat{a}^{\dagger 2}-\tilde{\kappa}_{+}^{*}\hat{a}^{2})
F(\hat{n})
\exp(\tilde{\kappa}_{-}\hat{a}^{\dagger 2}-\tilde{\kappa}_{-}^{*}\hat{a}^{2})
|m\rangle_{P},
\end{equation}
where $\tilde{\kappa}_{\pm}$ are given by Eq.~(\ref{tilde-kappa-plus-minus-definition-0}) and
\begin{eqnarray}
F(\hat{n})
&=&
\exp
\left[
iT^{2}|\kappa|^{2}
\frac{\tilde{T}_{+}-\sin\tilde{T}_{+}}{\tilde{T}_{+}}f(\hat{n})
\right]
e^{i\hat{H}_{0}^{(+)}T}
e^{-i\hat{H}_{0}^{(-)}T} \nonumber \\
&&
\times
\exp
\left[
-iT^{2}|\kappa|^{2}
\frac{\tilde{T}_{-}-\sin\tilde{T}_{-}}{\tilde{T}_{-}}f(\hat{n})
\right].
\end{eqnarray}
Thus, we obtain
\begin{equation}
{}_{P}\langle m|\hat{V}(t_{1},t_{0})|m\rangle_{P}
=
{\cal F}(m)+O(|\kappa|^{3}),
\end{equation}
where ${\cal F}(m)$ is given by Eq.~(\ref{cal-F}).

\section{\label{section-appendix-B}The exact solution of the time evolution with the Hamiltonian given by Eq.~(\ref{exactly-solvable-Hamiltonian-0})}
In this section, we derive the exact solution of the time evolution of the system whose Hamiltonian is given by Eq.~(\ref{exactly-solvable-Hamiltonian-0}).
By the method explained in Appendix~\ref{section-appendix-A}, we can derive
\begin{eqnarray}
\hat{V}'(t_{1},t_{0})
&=&
e^{i\hat{H}'_{+}(t_{1}-t_{0})}e^{-i\hat{H}'_{-}(t_{1}-t_{0})} \nonumber \\
&=&
\exp[2(\tilde{g}\hat{a}^{\dagger}-\tilde{g}^{*}\hat{a})] \nonumber \\
&=&
e^{2|\tilde{g}|^{2}}
\exp(-2\tilde{g}^{*}\hat{a})\exp(2\tilde{g}\hat{a}^{\dagger}),
\end{eqnarray}
where
\begin{equation}
\hat{H}'_{\pm}
=
\omega\hat{a}^{\dagger}\hat{a}
\pm
(g\hat{a}^{\dagger}+g^{*}\hat{a}),
\end{equation}
\begin{equation}
\tilde{g}
=
\frac{T}{\tilde{T}}(e^{i\tilde{T}}-1)g,
\end{equation}
$\tilde{T}=T\omega$, and $T=t_{1}-t_{0}$.
Then, a matrix element is given by
\begin{eqnarray}
{}_{P}\langle m|\hat{V}'(t_{1},t_{0})|m\rangle_{P}
&=&
e^{2|\tilde{g}|^{2}}
\sum_{k=0}^{\infty}
{}_{P}\langle m|\frac{1}{(k!)^{2}}(-4|\tilde{g}|^{2})^{k}\hat{a}^{k}\hat{a}^{+ k}|m\rangle_{P} \nonumber \\
&=&
e^{2|\tilde{g}|^{2}}
\sum_{k=0}^{\infty}
\frac{1}{(k!)^{2}}
(-4|\tilde{g}|^{2})^{k}
\frac{(m+k)!}{m!} \nonumber \\
&=&
e^{2|\tilde{g}|^{2}}
{}_{1}F_{1}(1+m;1;-4|\tilde{g}|^{2}),
\end{eqnarray}
where ${}_{1}F_{1}(a;b;z)$ is the Kummer confluent hypergeometric function.
Substituting the matrix element for ${\cal F}(m)$ in Eq.~(\ref{Tr-hat-V-t1-t0-hat-rho-0-P}), we obtain $\mbox{Tr}[\hat{V}'(t_{1},t_{0})\hat{\rho}_{0,P}]$.
Then, replacing
\\
$\mbox{Tr}[\hat{V}(t_{1},t_{0})\hat{\rho}_{0,P}]$ with
$\mbox{Tr}[\hat{V}'(t_{1},t_{0})\hat{\rho}_{0,P}]$ in Eq.~(\ref{expectation-value-hat-A-t1-t0-0}), (\ref{expectation-value-hat-A-t1-to-hat-A}), and (\ref{expectation-value-hat-A-hat-A-t1-to-hat-A}), we can derive $\mbox{LG}_{1,-1}$ for the system whose Hamiltonian is given by Eq.~(\ref{exactly-solvable-Hamiltonian-0}).
In particular, for the zero-temperature case, that is, $\beta=+\infty$, we obtain
\begin{equation}
\mbox{LG}_{1,-1}
=
\frac{1}{4}
\left\{
2-\sqrt{2}
+
\sqrt{2}\cos(T\Omega)
\exp
\left[
\frac{4g^{2}(-1+\cos(T\omega))}{\omega^{2}}
\right]
\right\}.
\end{equation}

\end{document}